\documentclass{aa}  
\usepackage{graphicx}
\usepackage{txfonts}
\usepackage{lipsum}
\usepackage{subcaption}                              
\usepackage{lscape}                                         
\usepackage{placeins}          
\usepackage{xcolor}
   
\begin{document}
\title{A universal bilinear model for the dependence of void asymmetry distributions on $\Omega_{m}$ and $\sigma_{8}$}
\titlerunning{void asymmetry distribution}
\author{Geonwoo Kang\thanks{woo@snu.ac.kr} \and Jounghun Lee\thanks{jounghun@snu.ac.kr}}
\authorrunning{Kang \& Lee}
\institute{Astronomy Program, Department of Physics \& Astronomy, Seoul National University, Seoul 08826, Republic of Korea}
\date{Received March 17, 2026}
\abstract
 {}
 {We present a new independent diagnostics based on void properties, by which the matter density parameter ($\Omega_{m}$) and amplitude of initial density field 
($\sigma_{8}$) can be constrained. This diagnostics utilizes coherent rotation of void galaxies, which can be observed as redshift asymmetry 
in opposite sides bisected by the projected spin axes of hosting voids.} 
{Identifying the voids and their member galaxies in the AbacusSummit of cosmological simulations and searching for their projected spin axes with an iterative method, 
we numerically determine the void asymmetry distributions both in real and redshift spaces for three different classes of the background 
cosmologies:  the $\Lambda$CDM, the $w$CDM and the $w_{0}w_{a}$CDM.}
{It is discovered that the void asymmetry distributions in real and redshift spaces are well approximated by the generalized Gamma models 
characterized by the scale and shape parameters for all of the $33$ background cosmologies considered. 
It turns out that the initial conditions affect only the scale parameter of the void asymmetry distributions that exhibits an almost linear dependence on each of 
$\Omega_{m}$ and $\sigma_{8}$ with being insensitive to $w$. 
Developing a bilinear model for the dependence of void asymmetry distributions on $\Omega_{m}$ and $\sigma_{8}$, we show that its coefficients are universal 
constants over the three cosmological classes. }
{Given that the void asymmetry distributions in redshift space are readily measurable properties, it is concluded that our universal model for the scale parameter of the void 
asymmetry distributions will contribute to breaking the $\Omega_{m}$-$\sigma_{8}$ degeneracy, regardless of $w$, when combined with 
the standard large-scale structure diagnostics.}


\maketitle
\nolinenumbers

\section{Introduction}\label{sec:intro}

The cosmic web is a jargon coined to describe a strikingly anisotropic interconnectivity that the galaxies exhibit in their spatial distributions when 
viewed on the largest-scale \citep{1996Natur.380..603B}. It has four distinct segments: knots, filaments, sheets and voids, often classified by 
the signs of the eigenvalues of local tidal tensors based on the Zel'dovich approximation~\citep{1970A&A.....5...84Z,1996ApJ...470L..41K} and 
characterized by their dynamical stabilities~\citep{2007MNRAS.375..489H}. 
The first structures ever formed via gravitational collapse of density inhomogeneities are the one-dimensional (1D) sheets, sometimes called 
the Zel'dovich pancakes~\citep{1970A&A.....5...84Z,1996ApJ...470L..41K}. 
Meanwhile, the most anisotropic structures hosting the largest galaxy population are the two-dimensional (2D) filaments~\citep{2019MNRAS.487.1607G}. 
At the junctions of multiple filaments are located the zero-dimensional (0D) knots where the largest bound objects, galaxy clusters, reside. 
These three over-dense segments have been the focus of extensive  studies to understand the origin and evolution of the cosmic web 
and to find its connection to the initial conditions of the universe
~\citep[e.g.,][and references therein]{1995MNRAS.272..231P,1996Natur.380..603B,2004MNRAS.353..162S,2008MNRAS.383.1655S,2009MNRAS.397.2163P,2014MNRAS.441.2923C,2021NatAs...5..839W,2021MNRAS.507.2968W,2023JCAP...02..058F,2024A&A...684A..63G,2024MNRAS.52711256L,2025MNRAS.539..873F}.

Recently, growing cosmological interest has been directed toward the fourth type of the cosmic web, the voids, 
markedly three dimensional (3D) structures wrapped by filaments and sheets, occupying the largest volume of the universe, but almost 
devoid of galaxies~\citep{1982Natur.300..407Z,1984MNRAS.206P...1I,1988ARA&A..26..245R,1996ApJ...470..160R,1997ApJ...491..421E,2005MNRAS.360..216C,2019MNRAS.487.1607G}.  
The voids are believed to be most vulnerable to the large-scale tidal effects due to their lowest-densities even at the primordial stages~\citep{2007PhRvL..98h1301P}
and to retain well the imprints of early universe physics due to their pristine nature~\citep{2002ApJ...566..641H, 2009MNRAS.397.2163P, 2023Natur.619..269D}.
Much effort has been made to find optimal void statistics that have the power to put tight constraints on the initial conditions 
of the universe~\citep[e.g.,][]{2004MNRAS.350..517S,2007PhRvL..98h1301P,2009ApJ...696L..10L,2010PhRvD..82b3002B,2012MNRAS.426..440B,2013MNRAS.434.1192R,2015JCAP...11..018M,2019JCAP...12..040V,2020JCAP...12..023H,2020ApJ...902..102R,2021MNRAS.507.2267D,2025A&A...695A..19F}. 

Very recently, the void spin distribution has been proposed by \citet{2025JCAP...06..011K} as a probe of the amplitude of initial density power spectrum, $\sigma_{8}$ 
and the matter density parameter, $\Omega_{m}$. 
Despite that the voids expand faster than the rest of the universe, their constituents develop tangential velocities as well as anisotropic spatial distributions 
within the voids due to the large-scale tidal effects~\citep{2006MNRAS.367.1629S}. 
It was \citet{2006ApJ...652....1L} who first introduced the concept of void spins to effectively describe the deviation of void galaxies from isotropic spatial distributions and 
radial motions.   Defining the angular momentum vectors of voids in the same manner as those of DM halos,  
\citet{2006ApJ...652....1L} showed that the directions of void spin vectors were strongly aligned with the 
principal axes of the largest-scale tidal fields, revealing strong connections between the voids and the linear tidal fields for the case of the 
standard cosmology where the anti-gravitational cosmological constant, $\Lambda$, and the cold dark matter (CDM), dominantly fill a flat universe. 

Using the same definition of the halo spin parameters as in~\citet{2001ApJ...555..240B} for the magnitudes of void spins,  \citet{2025JCAP...06..011K} 
newly found with the help of high-resolution $N$-body experiments that the probability density of void spin magnitudes is very well approximated by the 
generalized Gamma distribution not only for the standard $\Lambda$CDM but also for the quintessence dark energy (DE) models. 
The generalized Gamma model for the void spin distribution is characterized by two adjustable parameters whose best-fit values turned out to be dependent strongly on 
$\sigma_{8}$ and weakly on $\Omega_{m}$, while being insensitive to the other initial conditions such as the expansion rate, total neutrino mass ($M_{\nu}$) 
and DE equation of state ($w$).

In practice, however, it is not so plausible to determine the void spin distribution in real space with high accuracy since it requires information on the 3D 
peculiar velocities of the void galaxies.  
To overcome this practical difficulty, \citet{2025JCAP...06..011K} proposed a scheme based on the novel methodology devised by 
\citet{2021NatAs...5..839W} to determine the filament spins in 2D redshift space. According to this scheme, the projected spin axis of a void, if not being 
parallel to the line of sight direction, would divide the projected void region into two sectors between which significant asymmetry in the void galaxy 
redshifts is expected to exist.  The projected spin axis of a void can be determined as the bisector line that yields the maximum redshift 
asymmetry between the two sides. The void spins, or equivalently the magnitudes of rotational motions of the void galaxies around the spin axes, 
should be proportional to this maximum redshift asymmetry. 

At this stage, three critical questions arise about this scheme: First, would the void asymmetry distribution measured in 2D redshift space via this scheme be also well described by the 
generalized Gamma model regardless of a background cosmology like the original 3D void spin distribution? Second, would it also show the same or at least similar degree of cosmology 
dependence?  Third, would it be practically feasible to constrain $\Omega_{m}$ and $\sigma_{8}$ with observed void asymmetry distributions?
In this Paper, we attempt to answer these three questions by creating and analyzing a mock redshift space dataset from a series of high-resolution $N$-body simulations 
performed for a comprehensive broad range of background cosmologies including dynamical DE models with time evolving $w$ values. 

\section{Physical analysis}\label{sec:analysis}

Our numerical analysis relies on the data from the AbacusSummit series of simulations~\citep{2021MNRAS.508.4017M} that were run by applying the ABACUS 
cosmological $N$-body code~\citep{2021MNRAS.508..575G} to $\sim 330$ billion DM particles with mass resolution of $2\times 10^{9}\,h^{-1}\,M_{\odot}$ 
on a periodic box of volume $8\,h^{-3}{\rm Gpc}^{3}$. The AbacusSummit simulations kept tracks of the DM particles from the initial redshift $z=99$ 
down to $z=0.1$ for the most extensive scope of background cosmologies, the initial conditions of which were all efficiently set up according 
to the Cosmic Linear Anisotropy Solving System (CLASS)~\citep{2011arXiv1104.2932L}.
From the particle snapshots at redshifts in the range of $0.1\le z\le 8$ were resolved the bound DM halos via the COMPASO halo-finder that 
\citet{2022MNRAS.509..501H} had developed by refining the classical spherical over-density algorithm~\citep{1992ApJ...399..405W}.

\begin{figure}[ht!]
   \centering
   \includegraphics[width=\hsize]{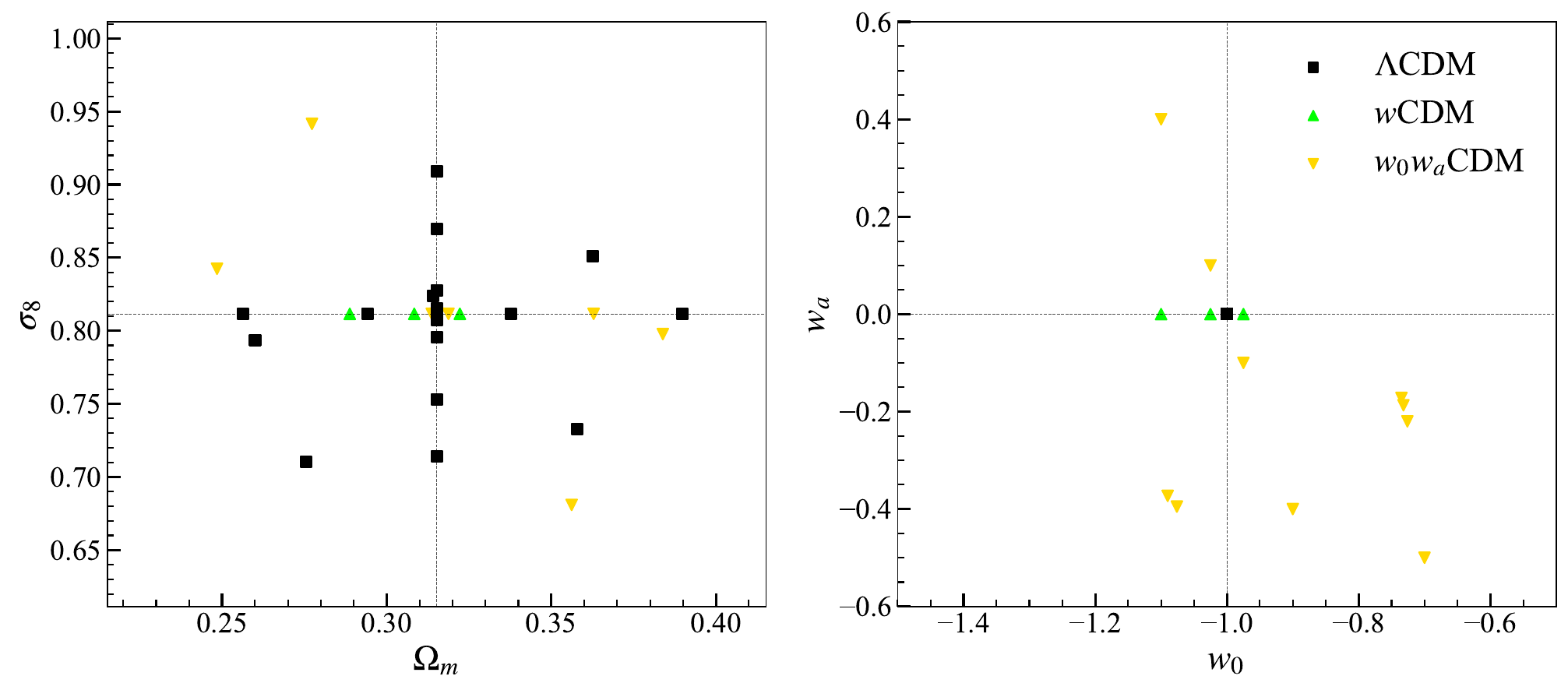}
      \caption{Configurations of $33$ different background cosmologies in 2D spaces spanned by 
$\Omega_{m}$-$\sigma_{8}$ (left panel) and by $w_{0}$-$w_{a}$ (right panel).}
         \label{fig:cfg}
   \end{figure}

Among the cosmologies covered by the AbacusSummit series are considered three different classes for the purpose of the current work. 
To the first class belongs the flat $\Lambda$CDM cosmologies with various values of $\Omega_{m}$ and $\sigma_{8}$, including 
the one with the latest Planck initial conditions~\citep{2020A&A...641A...6P}. The second class comprises the quintessence DE cosmologies with constant 
equation of state $w\ne -1$ and  $dw/da=0$ with scale factor $a$, while the third class embraces the dynamical DE models, $w_{0}w_{a}$CDM, with time varying equation 
of states, $w(a)=w_{0}+(1-a)w_{a}$, what is called, the Chevallier-Polarski-Linder parameterization~\citep{2001IJMPD..10..213C,2003PhRvL..90i1301L}. 
For the current analysis are selected a total of $33$ cosmologies ($20$, $3$ and $10$ from the $\Lambda$CDM, $w$CDM and $w_{0}w_{a}$CDM class, respectively). 
Fig.~\ref{fig:cfg} displays how the selected cosmologies of each class are distributed in the configuration space spanned by 
$\Omega_{m}$-$\sigma_{8}$ (left panel) and by $w_{0}$-$w_{a}$ (right panel). 

For each cosmology, we select only those DM halos with total masses $M/(h^{-1}\,M_{\odot}) \ge 10^{11.5}$ at $z=0.1$, the lowest redshift available from the AbacusSummit simulations, 
to create a sample of galactic halos at the present epoch. Hereafter, we will interchangeably use two terms, galactic halos and galaxies, to refer to the selected DM halos. 
For the determination of the mass cutoff, $\log\left[M_{\rm cut}/(h^{-1}\,M_{\odot})\right]=11.5$, we consider the Vera C. Rubin Observatory 
Legacy Survey of Space and Time (LSST) by the LSST Dark Energy Science Collaboration (LSST DESC)~\citep{2009arXiv0912.0201L} as a reference survey. 
We analyze the photometric galaxy catalog from the simulated sky survey data preview 2.0 of the LSST DESC~\citep{2021ApJS..253...31L} in the redshift range of $0.1\le z\le 0.2$ 
that was obtained by mimicking a single exposure observation with elaborate photo-$z$ algorithm. Then, we extract the halo masses of individual galaxies whose $g$-band magnitudes 
exceed the flux-limit set by  the LSST DESC to determine the minimum mass,  $M_{\rm cut}$, beyond which the largest number of the galaxies reside.

To be consistent with real observations where the void spins can be found only in the plane perpendicular to the line-of-sight (los) direction, 
we create a mock redshift space galaxy sample.  Using the flat-sky approximation and under the assumption that the los direction is parallel to the $\hat{x}_{3}$-axis, 
the los component of the redshift space position of a galaxy located at $(x_{1},x_{2},x_{3})$ in real space is determined as 
$x_{r3}\equiv x_{3}+{\bf v}\cdot \hat{x}_{3}/H_{0}$ where ${\bf v}$ is the peculiar velocity and $H_{0}$ is the Hubble constant given as $H_{0}\equiv 100\,h\,{\rm km}\,s^{-1}\,{\rm Mpc}^{-1}$ 
\citep{1998ASSL..231.....H}.  The other two components of the redshift space position is the same as the real space counterpart.

To be consistent with \citet{2025JCAP...06..011K}, we adopt the classical Void-Finder algorithm~\citep{2002ApJ...566..641H} and apply it 
to this mock redshift space galaxy sample.  Basically, this algorithm identifies a void as a union of a maximal sphere and non-maximal spheres intersecting the maximal one, 
which fit the volume of an empty region containing no wall galactic halos. 
The volume of each void, $V_{\rm v}$, is computed with the help of the Monte-Carlo method and then its effective radius, $R_{\rm v}$, is determined  
as $R_{\rm v}\equiv [3V_{\rm v}/(4\pi)]^{1/3}$~\citep{2006ApJ...652....1L,2007PhRvL..98h1301P}. 
For the detailed description of the classification of the wall and field galactic halos as well as the void-identification procedure via the Void-Finder algorithm, 
we refer the readers to ~\citet{2002ApJ...566..641H}, \citet{2006ApJ...652....1L} and \citet{2025JCAP...06..011K}.

Among the identified voids, we select only those voids within which $15$ or more field galaxies above the mass-cut are embedded. 
If a void is truly spinning, the projection of its spin axis onto the plane of sky will bisect the void into two sides between which a significant difference exists 
in the relative peculiar velocities along the los direction. One of the two sectors possesses the galaxies whose residual velocities in the direction 
of $ \hat{x}_{3}$ relative to the center will be in the direction of $\hat{x}_{3}$, while the galaxies in the other sector will appear to move in the direction of 
$-\hat{x}_{3}$. Hereafter, a net difference between the rescaled los velocities of the galactic halos belonging to the two opposite sectors, 
will be called the redshift asymmetry, $\Delta z \equiv \Delta v/c$.

In real observations, the orientations of the projected spin axes of voids are unlikely to be directly attainable, since it requires information on the 3D 
peculiar velocities, the measurements of which are greatly hindered by the practical difficulty in determining the peculiar velocities of the void galaxies. 
Notwithstanding,  the orientation of the projected spin axes of voids may be effectively found by employing the same methodology used in the seminal 
work of \citet{2021NatAs...5..839W} who determined the 2D orientations of filament spin axes in the plane of sky. 
We randomly generate an orientation angle of a bisector from the $x_{1}$-axis in the $x_{1}$-$x_{2}$ plane, passing through the center of a void. 
Then, we calculate the redshift asymmetry between the opposite sides of the bisector.  Iterating this process until the randomly generated orientation angle 
spans a full range from $0$ to $180^{\circ}$, we inspect at what orientation angle the redshift asymmetry reaches the maximum value, 
$\Delta z_{\rm max}$. 

What follows is a detailed description of the procedure through which the redshift asymmetry of a void is determined: 
Let ${\bf x}_{\beta}^{2d}$ denote the comoving position vector of the $\beta$-th galaxy belonging 
to a void consisting of $N$ members, located at the center position, ${\bf x}_{c}^{2d}$, in 2D redshift space normal to the los direction, and 
let ${\bf v}_{\beta}^{1d}$ be the los component of its peculiar velocity relative to the void center. Let also $\hat{\bf d}$ denote the 2D unit vector parallel to a bisector 
passing through ${\bf x}_{c}^{2d}$.  The redshift asymmetry in the opposite sectors dichotomized by $\hat{\bf d}$ is computed as 
\begin{equation}
\label{eqn:dz}
\Delta z = \frac{1}{c}\Bigg{\vert}\sum_{\beta=1}^{N}{}{\sigma_\beta {\bf v}_{\beta}^{1d}} \Bigg{\vert}\, ,
\end{equation}
where $\sigma_{\beta}\equiv {\rm sgn}\left[\hat{\bf d}\times ({\bf x}_{\beta}^{2d}-{\bf x}_{c}^{2d})\right]$. Note that if $\sigma_{\beta}>0$ ($\sigma_{\beta}<0$), 
then the $\beta$-th void galaxy belongs to the counterclockwise (clockwise) sector in the face-on image of its hosting void \citep{2021NatAs...5..839W}. 
We search for the orientation of a bisector, $\hat{\bf d}$, that maximizes $\Delta z$, and determine it as the projected spin axis of a given void. 

It is worth stressing that ${\bf v}_{\beta}^{1d}$ in Eq.~(\ref{eqn:dz}) is an observable quantity, which will be obtained in practice from the los 
(i.e., radial) velocity of a given void galaxy (or equivalently from the measurement of its redshift residual) that includes both of the Hubble flow and peculiar velocity terms. 
In the current numerical work, however, where the mock galaxy catalog is obtained from the snapshot of particle distributions at a fixed redshift rather than from the 
light-cone distributions over a broad redshift range, the contribution of the Hubble flow to the los radial velocity vanishes, which is why the los radial 
velocity is equivalent to the los component of the peculiar velocity in Eq.~(\ref{eqn:dz}).

   \begin{figure}[ht!]
   \centering
   \includegraphics[width=\hsize]{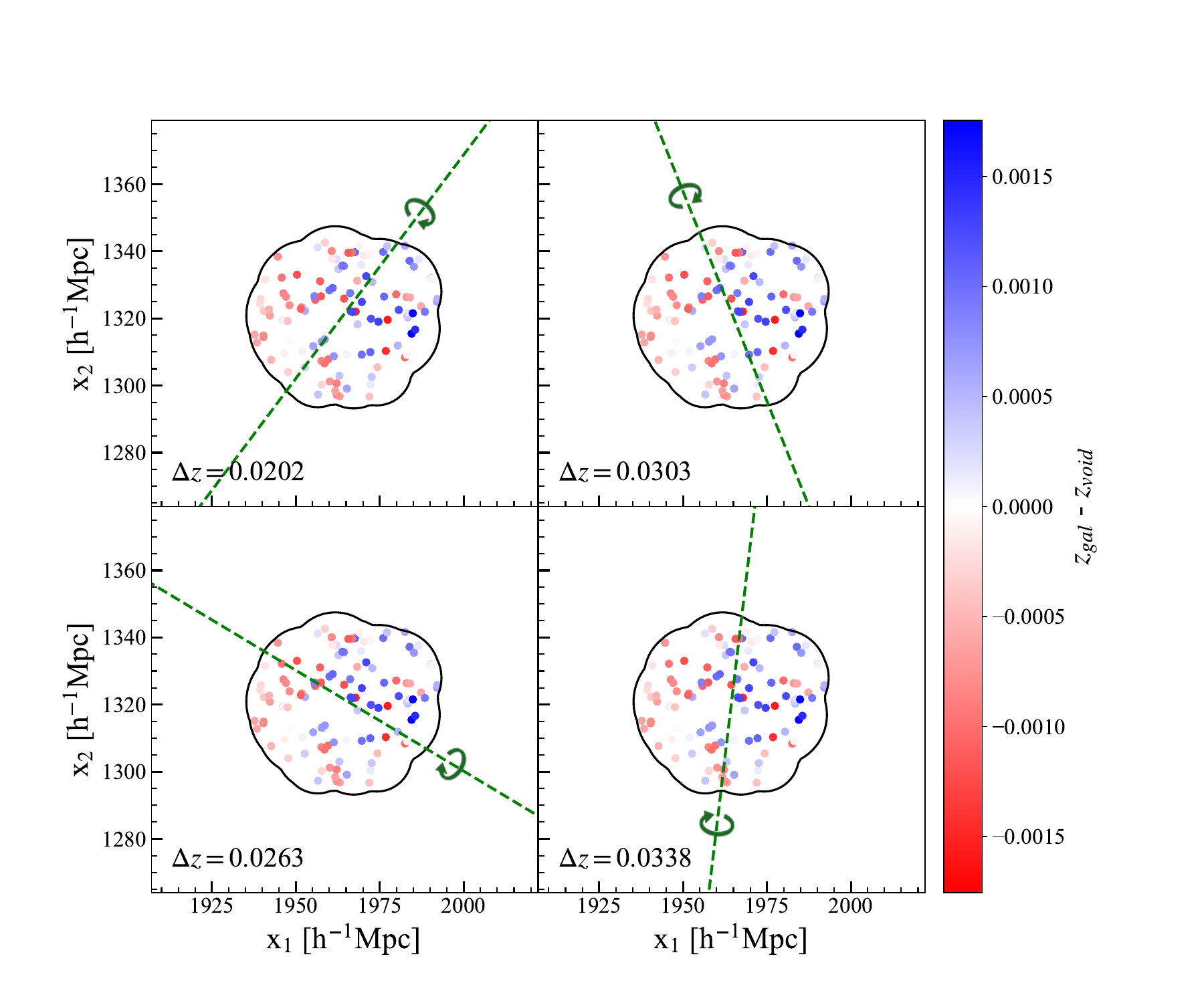}
      \caption{Illustration of the redshift asymmetries of void galaxies for four different cases of the orientation of the bisector line. In each panel,  the green dashed line represents 
      the bisector, each filled circle corresponds to a void galaxy in the 2D space normal to the los direction, the color of which reflects its redshift residual relative to the void center, while $\Delta z$ value 
      indicates the redshift asymmetry of a given void computed by Eq.~(\ref{eqn:dz}). 
      The axis of void rotation projected onto the plane perpendicular to the los direction is found as the bisector line about which the largest redshift 
      asymmetry of void galaxies is witnessed (bottom-right panel). }
         \label{fig:ill}
   \end{figure}
Fig.~\ref{fig:ill} illustrates a projected giant void in the $x_{1}$-$x_{2}$ plane, depicting how the redshift asymmetry of its member galaxies changes 
as the orientation of the bisector (dashed green line) alters.  The bisector that yields the maximum redshift asymmetry, $\Delta z_{\rm max}$,  
corresponds to the projected 2D spin axis of this void (bottom-right panel). This maximum value of $\Delta z_{\rm max}$ reflects how rapidly and coherently 
the member galaxies rotate around the spin axis.  Determining the values of $\Delta z_{\rm max}$ of all voids for each cosmology, we numerically obtain the 
probability density distribution, $p(\Delta z_{\rm max})$, by dividing the full range of $\Delta z_{\rm max}$ into short bins and counting the numbers of voids 
whose values of $\Delta z_{\rm max}$ fall in each bin. The associated uncertainty in $p(\Delta z_{\rm max})$ at each bin is computed as the Poisson error. 
From here on, we refer to $p(\Delta z_{\rm max})$ as the void asymmetry distribution.

For the investigation of the net effect of the initial conditions on the void asymmetry distributions, however, it is important to use the controlled samples of 
voids whose size distributions (i.e., the number fractions of voids as a function of their sizes) are identical among different cosmologies. 
Given that the void size distributions depend on the initial conditions~\citep{2004MNRAS.350..517S,2019JCAP...12..040V} 
and that the magnitudes of void spins depend on their sizes \citep{2025JCAP...06..011K},  we suspect that a spurious signal of correlation between the initial conditions 
and $\Delta z_{\rm max}$ could be produced by the differences among the selected cosmologies in the void size distributions. 
To eliminate these possible secondary effects of initial conditions on the void asymmetry distributions, we create $R_{\rm v}$-controlled void samples 
by dividing the ranges of $R_{\rm v}$ into short intervals and drawing a specific number, $n_{v}$, of voids at each interval to make all of the $33$ cosmologies share the 
identical number fraction, $n_{v}/N_{\rm total}$, where $N_{\rm total}$ is the total number of selected voids for each cosmology. 

\begin{figure}[ht!]
   \centering
   \includegraphics[width=\hsize]{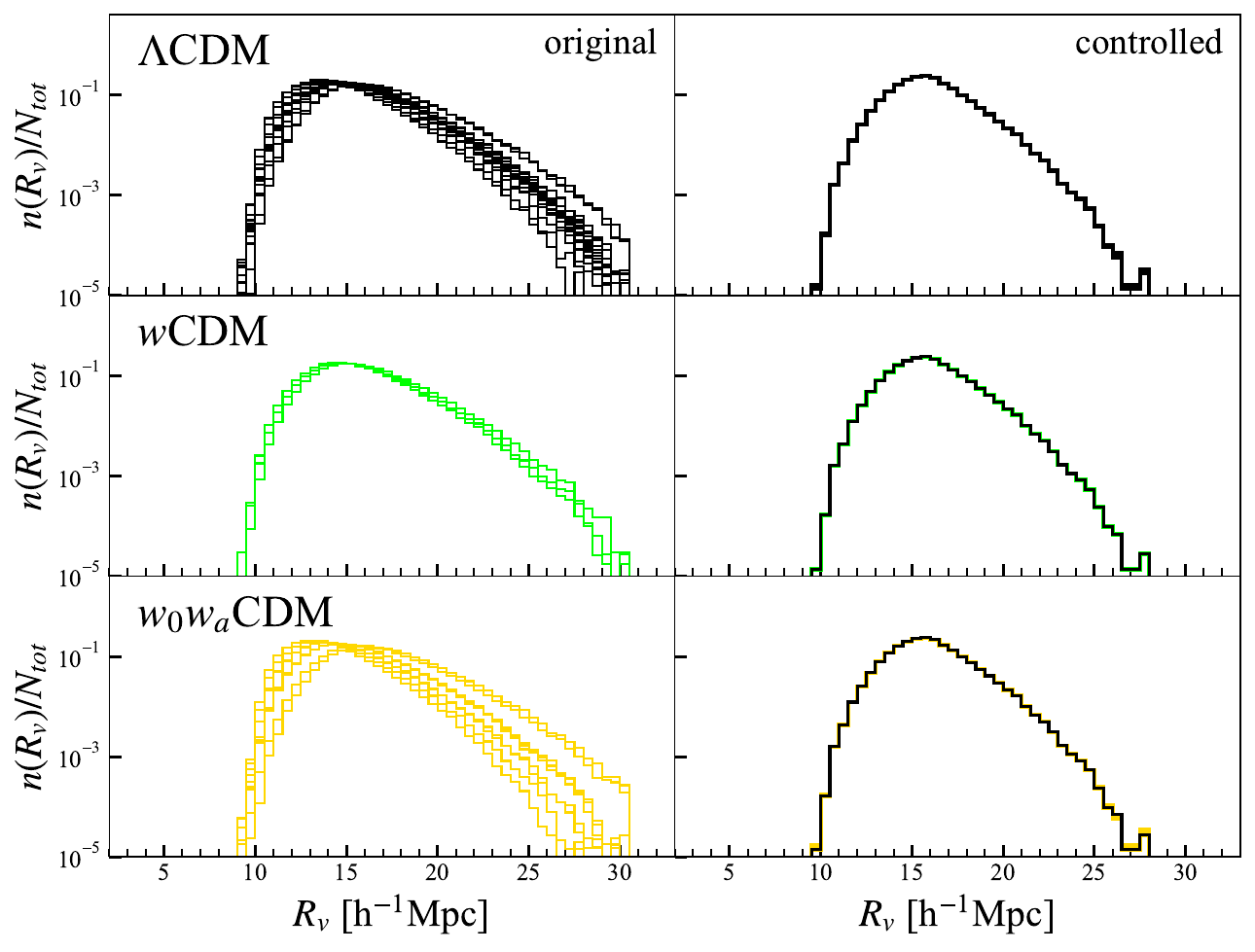}
      \caption{Number fractions of voids versus their sizes from the original and $R_{v}$-controlled samples (left and right panels, respectively) for the three classes of the background cosmologies. Note that the 
      controlled void size distributions are identical over the three classes.}
         \label{fig:nrv}
   \end{figure}

\begin{figure}[ht!]
   \centering
   \includegraphics[width=\hsize]{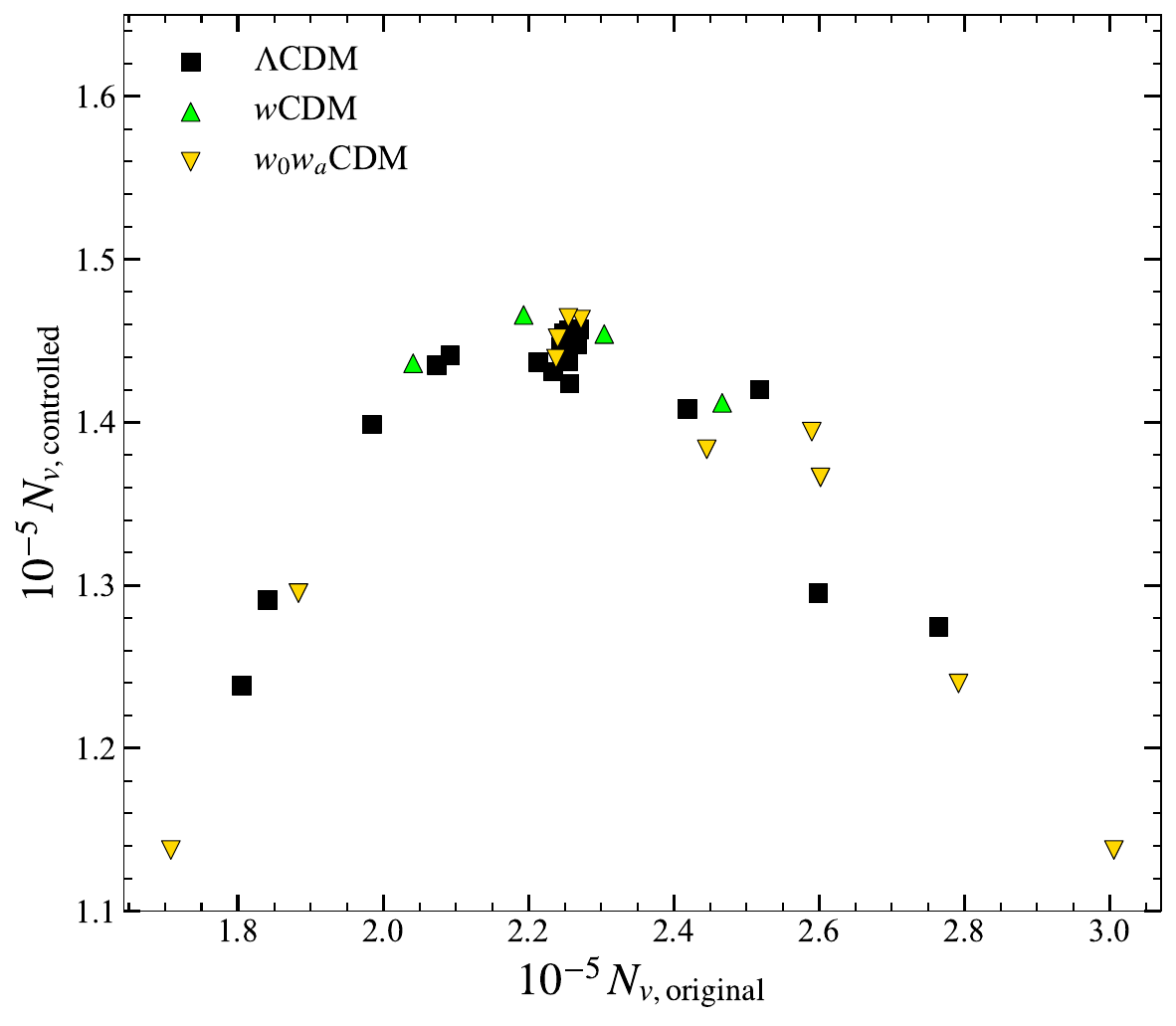}
      \caption{Numbers of voids having $15$ or more galaxies from the controlled vs. original samples for the $33$ cosmologies ($20$ $\Lambda$CDM, $3$ $w$CDM and 
      $10$ $w_{0}w_{a}$CDM cosmologies).}
         \label{fig:nv}
   \end{figure}

Fig.~\ref{fig:nrv} plots the number fraction of voids as a function of $R_{\rm v}$ from the original and controlled samples of voids (left and right panels, respectively). 
As can be seen, the $R_{\rm v}$-controlled samples yield the identical $n_{v}/N_{\rm total}$ distributions among all of the cosmologies considered, while the voids from the original samples 
have markedly different distributions from one another. 
If the void asymmetry distributions from the $R_{v}$-controlled samples show a significant signal of cosmology dependence, then it is now guaranteed that the signal is not due to  
any difference in void size distributions among the samples. Fig.\ref{fig:nv} shows the numbers of voids from  the $R_{v}$ controlled samples versus those from the original ones 
for all of the $33$ cosmologies. 

Recalling the result of ~\citet{2025JCAP...06..011K} that the probability density distribution of void spins determined in the 3D real space was well described by the 
generalized Gamma model regardless of the background cosmology, we also compare the numerically obtained $p(\Delta z_{\rm max})$ from the $R_{v}$-controlled samples 
with the same model: 
\begin{equation}
\label{eqn:gam}
p(\Delta z_{\rm max}) = \frac{(\Delta z_{\rm max})^{k-1}}{2\Gamma\left(2k\right)\theta^{k}}
\exp\left[-\left(\frac{\Delta z_{\rm max}}{\theta}\right)^{1/2}\right] \, ,
\end{equation}
where $\Gamma$ denotes the Gamma function, while $k$ and $\theta$ are two characteristic parameters, quantifying its shape and scale. 
The best-fit values of $k$ and $\theta$ are determined by fitting Eq. (\ref{eqn:gam}) to the numerically obtained $p(\Delta z_{\rm max})$ via the 
$\chi^{2}$-minimization process. 

   \begin{figure}[ht!]
   \centering
   \includegraphics[width=\hsize]{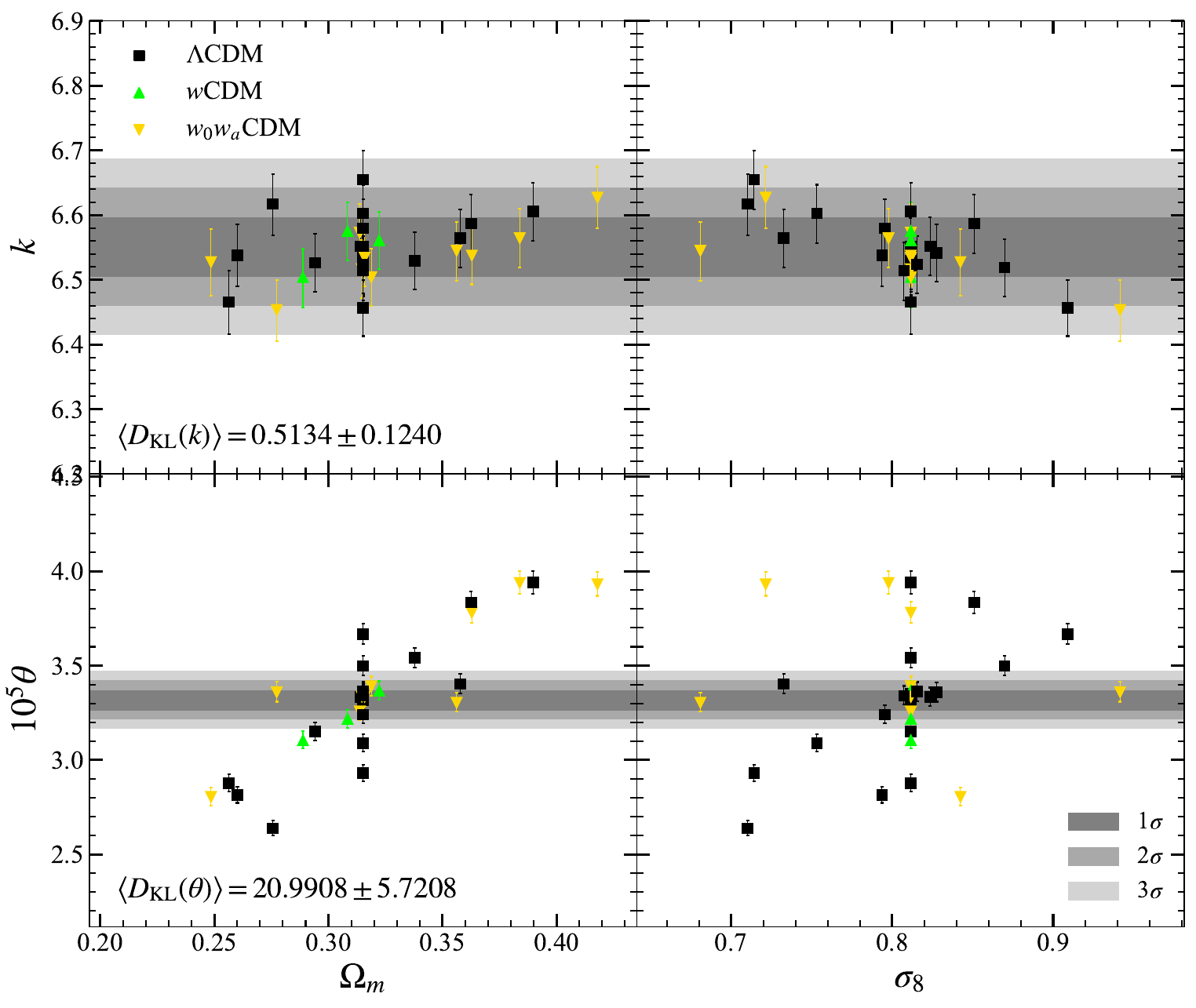}
      \caption{$\Omega_{m}$ and $\sigma_{8}$ variations (left and right panels, respectively) of two parameters, $k$ and $\theta$, of the generalized $\Gamma$-model for the 
      redshift asymmetry distributions of redshift space voids.  The mean KL divergences, $D_{\rm KL}$, are measured to quantify the sensitivity of each of $k$ and $\theta$ to the change of 
      $(\Omega_{m},\sigma_{8})$ from the Planck to non-Planck values.}
         \label{fig:kt}
   \end{figure}
Fig.~\ref{fig:kt} shows how the best-fit values of $k$ and $\theta$ vary with $\Omega_{m}$ and $\sigma_{8}$ of the background cosmologies considered. 
The error, $\sigma$, involved in the determination of each parameter corresponds to the marginalized one standard deviation computed under the assumption 
of Gaussianity: $p\left[\chi^{2}(k,\theta)\right] = \exp\left[-\chi^{2}(k,\theta)/2\right]$. 
The horizontal shaded dark, light and lightest gray regions correspond to the $1\sigma$, $2\sigma$ and $3\sigma$ confidence regions around the 
best-fit values of $k$ and $\theta$ for the case of the Planck $\Lambda$CDM cosmology~\citep{2020A&A...641A...6P}.  
As can be seen, the scale parameter, $\theta$, exhibits a sensitive variation with $\Omega_{m}$ and $\sigma_{8}$, while the shape parameter, $k$, seems 
to be almost independent of $\Omega_{m}$ and $\sigma_{8}$.  

To quantify how sensitively $(k, \theta)$ depends on $(\Omega_{m},\sigma_{8})$, we compute the Kullback-Leibler divergence, $D_{\rm KL}$ \citep{Kullback:1951zyt}:
\begin{equation}
D_{\rm KL}(s) = \int\, p_{\rm npl}(s)\ln\!\left[\frac{p_{\rm npl}(s)}{p_{\rm pl}(s)}\right]ds , \quad s\in \{k,\theta\} ,
\label{eqn:KL}
\end{equation} 
where $p_{\rm pl}(s)$ and $p_{\rm npl}(s)$ represent the Gaussian probability density functions of $s$ for the Planck and a non-Planck cosmology, respectively. 
The mean and standard deviation of each parameter are set at the best-fit value and its uncertainty shown in Fig.~\ref{fig:kt}. 
Basically, this KL divergence measures how strongly $(k,\theta)$ is affected by the change of $(\Omega_{m},\sigma_{8})$ from the Planck to non-Planck values.  
The larger $D_{\rm KL}(s)$ is, the more sensitive $s$ is to $(\Omega_{m},\sigma_{8})$. The mean values of $D_{\rm KL}(k)$ and $D_{\rm KL}(\theta)$ 
averaged over the $32$ non-Planck cosmologies are listed in Fig.\ref{fig:kt}.  As can be seen, the value of $\langle D_{\rm KL}(k)\rangle$ are negligibly small compared with 
those of $\langle D_{\rm KL}(\theta)\rangle$, which  confirms that only the scale parameter, $\theta$, of $p(\Delta z_{\rm max})$ is a sensitive probe of the background cosmology. 

\begin{figure}[ht!]
\centering
\includegraphics[width=\hsize]{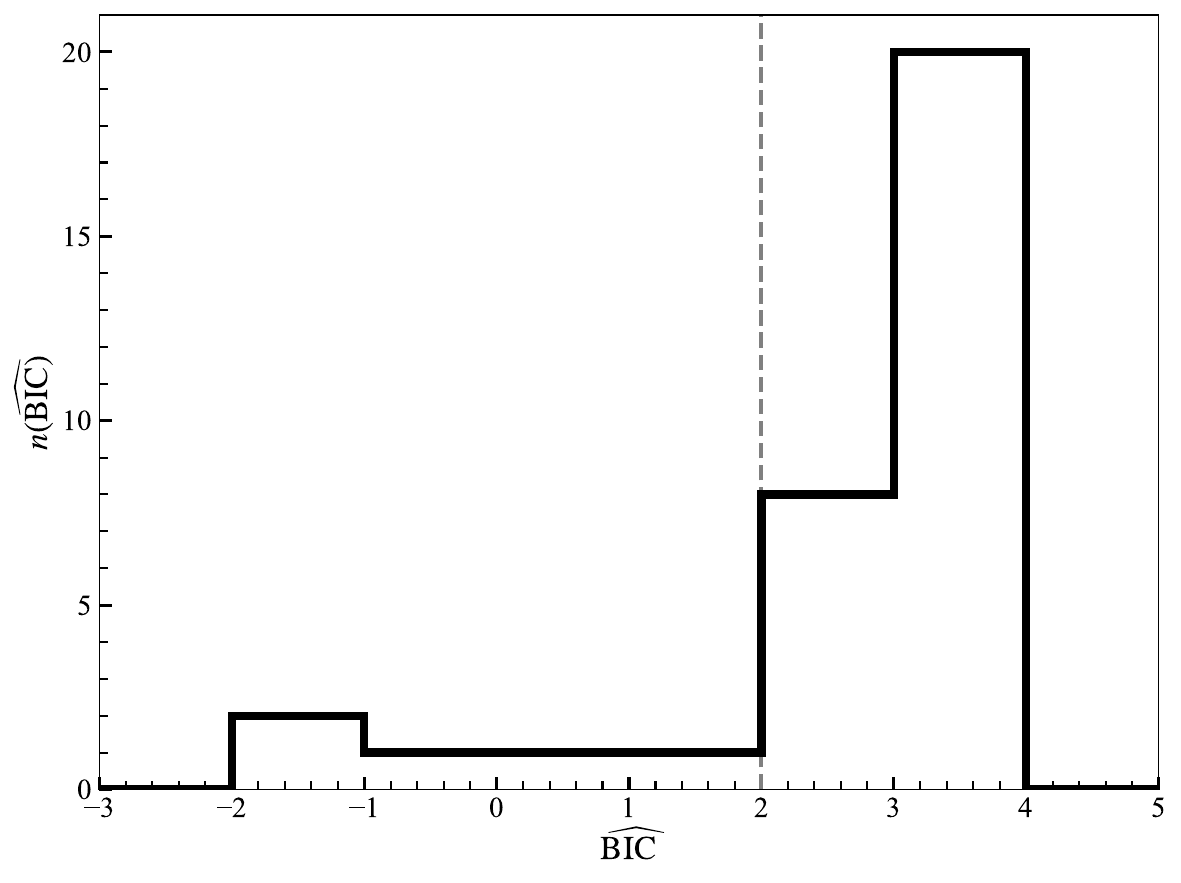}
\caption{Numbers of the background cosmologies vs. the $\widehat{\rm BIC}$ values evaluated by Eq.(\ref{eqn:bic}) between two models: 
The first model treats Eq.(\ref{eqn:gam}) as a single-parameter distribution, while the second model regards it as a double-parameter formula. 
For the background cosmologies whose $\widehat{\rm BIC}$ values exceed the threshold value of $2$ (vertical dashed line) \citep{kas95}, 
the first model may well be chosen as a superior one.}
\label{fig:bic}
\end{figure}

Upon this finding, we set the shape parameter at the fixed best-fit value, $k=6.551$, obtained for the Planck $\Lambda$CDM case, and treat Eq.~(\ref{eqn:gam}) 
as a single-parameter formula.  To justify this choice, we compute the Bayesian information criterion, $\widehat{\rm BIC}$, defined as \citep{Schwarz:1978tpv}
\begin{equation}
\widehat{\rm BIC} = 2\ln\left[\frac{\hat{L}_{1}}{\hat{L}_{2}}\right] - (m_{1}-m_{2})\ln n\, ,
\label{eqn:bic}
\end{equation} 
where $\hat{L}_{1}$ and $\hat{L}_{2}$ denote the maximum likelihoods of two models which have $m_{1}$ and $m_{2}$ parameters, respectively, both of 
which are fitted to $n$ data points, $\{\Delta z_{\rm max, j}, p(\Delta z_{\rm max,j})\}_{j=1}^{n}$ from the AbacusSummit simulation dataset for each cosmology.  
Both of the models assume that the numerically obtained data points follow the same generalized Gamma distribution, Eq.(\ref{eqn:gam}). But, they yield different maximum likelihood values 
(i.e., $\hat{L}_{1}\ne \hat{L}_{2}$) since the model 1 has only one free parameter, $\theta$ (i.e., $m_{1}=1$), while the model 2 has two free parameter, $\{k,\theta\}$ (i.e., $m_{2}=2$).  
It is found that the majority of the background cosmologies considered ($28$ out of $33$) yields $\widehat{\rm BIC}$ that exceeds the conventional threshold value of $2$ 
(Fig. \ref{fig:bic}) \citep{kas95}, which justifies our choice of treating Eq.(\ref{eqn:gam}) as a single-parameter model for $p(\Delta z_{\rm max})$.

\begin{figure}[ht!]
   \centering
   \includegraphics[width=\hsize]{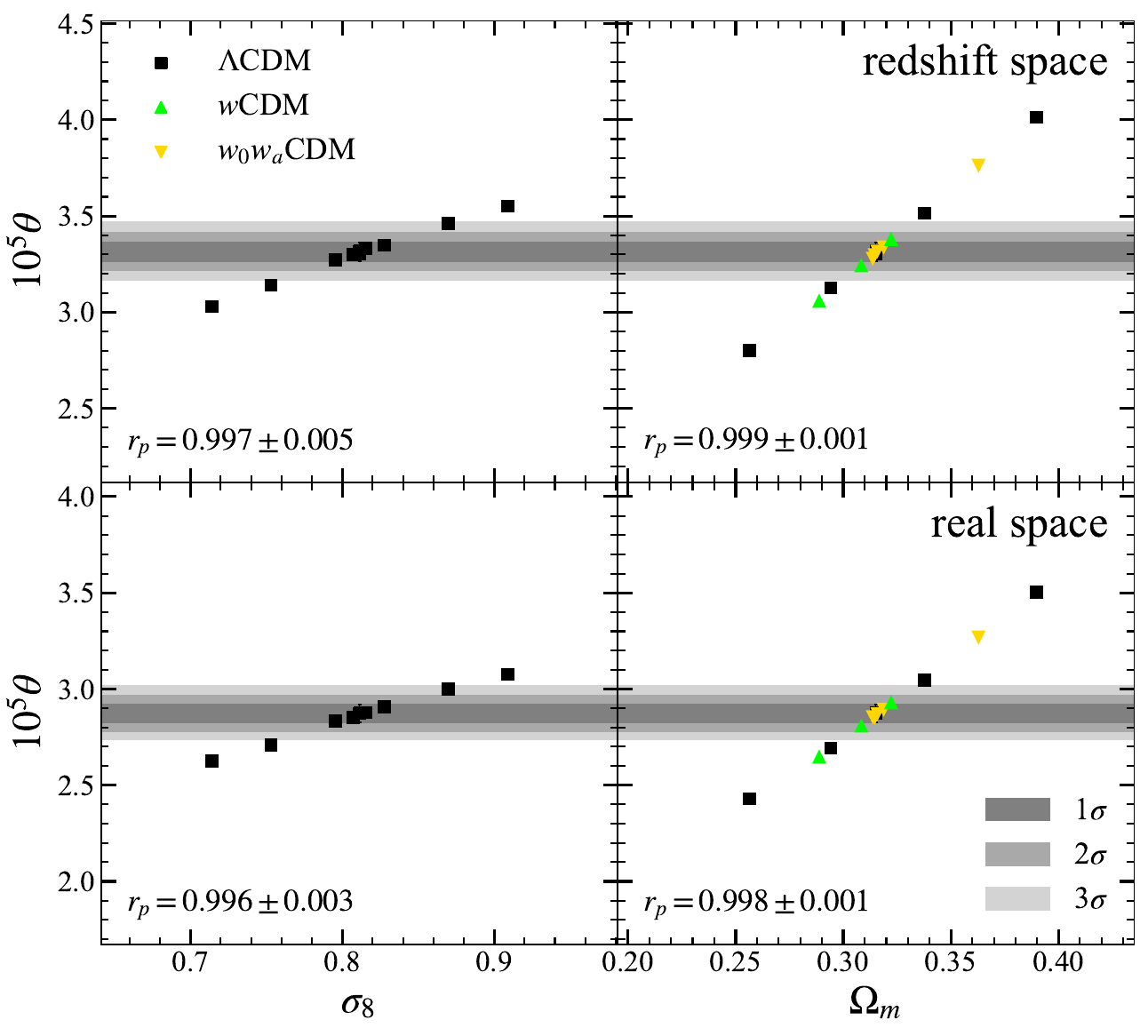}
      \caption{Variations of $\theta$, with each of $\Omega_{m}$ and $\sigma_{8}$ when one of the two cosmological parameters are fixed at the Planck values 
      in real and redshift space (top and bottom panels, respectively). 
      In each panel $r_{p}$ is the Pearson correlation coefficient that measures the linearity between $\theta$ and each of $\Omega_{m}$ and $\sigma_{8}$.}
         \label{fig:t_fix}
   \end{figure}

Now that only the scale parameter $\theta$ is confirmed to be sensitive to $\Omega_{m}$ and $\sigma_{8}$, we examine how the best-fit value 
of $\theta$ varies with each of $\sigma_{8}$ and $\Omega_{m}$, when one of the initial conditions is fixed at the Planck value. 
Among the $33$ cosmologies considered, we select those $15$ cosmologies ($10$, $3$, and $2$ from the $\Lambda$CDM, 
$w$CDM and $w_{0}w_{a}$CDM classes, respectively) for which $\Omega_{m}$ is set at the Planck value. 
We fit Eq.~(\ref{eqn:gam}) to the numerically obtained $p(\Delta z_{\rm max})$ for each of these cosmologies to re-determine the best-fit $\theta$, 
which is shown in the top-left panel of Fig.~\ref{fig:t_fix}. 
To quantify how linear the mutual dependence between $\theta$ and $\sigma_{8}$ is , we evaluate the Pearson correlation coefficient, $r_{p}$, over the $15$ cosmologies, 
which is also listed in the top-left panel of Fig.~\ref{fig:t_fix}.  As can be seen, the best-fit values of $\theta$ indeed show almost perfect linear dependence on $\sigma_{8}$, 
yielding $r_{p}\sim 1.0$. In a similar manner, we also investigate how $\theta$ changes with $\Omega_{m}$ at fixed $\sigma_{8}$ by determining the best-fit $\theta$ from the 
$11$ $\Lambda$CDM cosmologies for which $\sigma_{8}$ is set at the Planck value. An almost perfect linear dependence with $r_{p}\approx 1$ is also witnessed between 
$\theta$ and $\Omega_{m}$, as shown in the top-right panel of Fig.~\ref{fig:t_fix}.
\begin{figure}[ht!]
\centering
\includegraphics[width=\hsize]{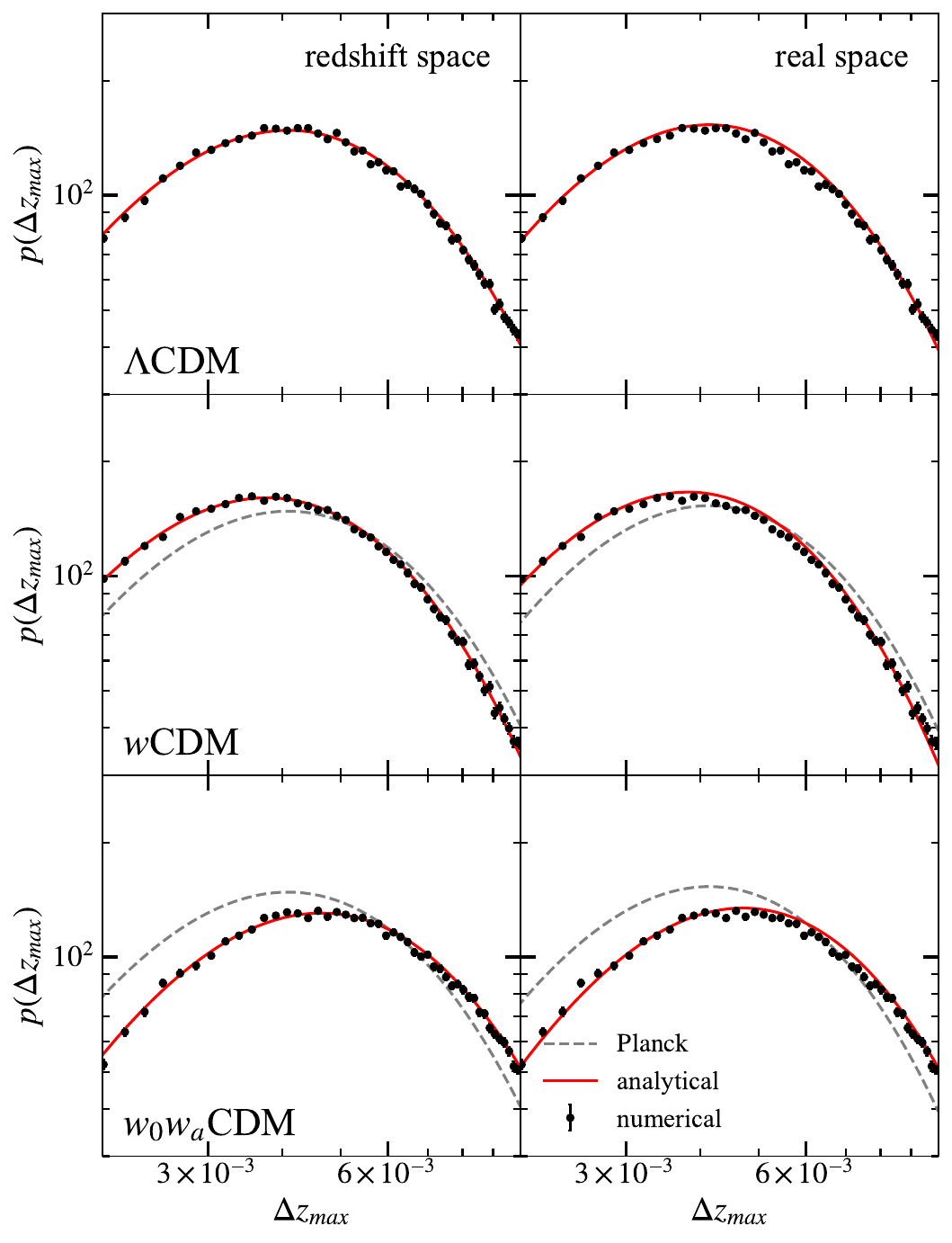}
\caption{Numerically obtained probability density distribution of void redshift asymmetry (filled black circles) compared with the best-fit 
$\Gamma$-distributions (red solid line) for three exemplary cosmologies: a $\Lambda$CDM cosmology with $\Omega_{m}=0.315$ and $\sigma_{8}=0.811$; 
a $w$CDM cosmology with $\Omega_{m}=0.289$, $\sigma_{8}=0.812$ and $w=-1.1$; a $w_{0}w_{a}$CDM cosmology 
with $\Omega_{m}=0.362$, $\sigma_{8}=0.812$, $w_{0}=-0.7$ and $w_{a}=-0.5$.}
\label{fig:gam}
\end{figure}

Fig.~\ref{fig:gam} shows $p(\Delta z_{\rm max})$ for three exemplary cosmologies  (black filled circles) and compares them with the analytical Gamma 
model (red solid lines) given in Eq. ~(\ref{eqn:gam}) with the best-fit value of $\theta$.  The shape parameter, $k$, is set at the Planck best-fit value of $6.551$ 
for all of the three cases. Despite that we adjust only the scale parameter, $\theta$, of the generalized Gamma distribution, Eq.(\ref{eqn:gam}) is still 
in remarkably good agreements with the numerical results for all of the three cases. 
Although Fig.\ref{fig:gam} displays the results only for the three exemplary cosmologies, we confirm that the single 
parameter formula, Eq.(\ref{eqn:gam}), validly describes $p(\Delta z_{\rm max})$ for all of the $33$ cosmologies considered. 
Recalling the result of~\citet{2025JCAP...06..011K} that the probability density distribution of void spins determined in 3D real-space was well 
described by the same generalized Gamma distribution, we interpret these good agreements between $p(\Delta z_{\rm max})$ and Eq.(\ref{eqn:gam}) for 
all of the cosmologies considered as an evidence that the void asymmetry distributions measured in projected 2D redshift space indeed are an efficient 
practical proxy of the true void spin distributions defined in 3D real-space.

\begin{figure}[ht!]
   \centering
   \includegraphics[width=\hsize]{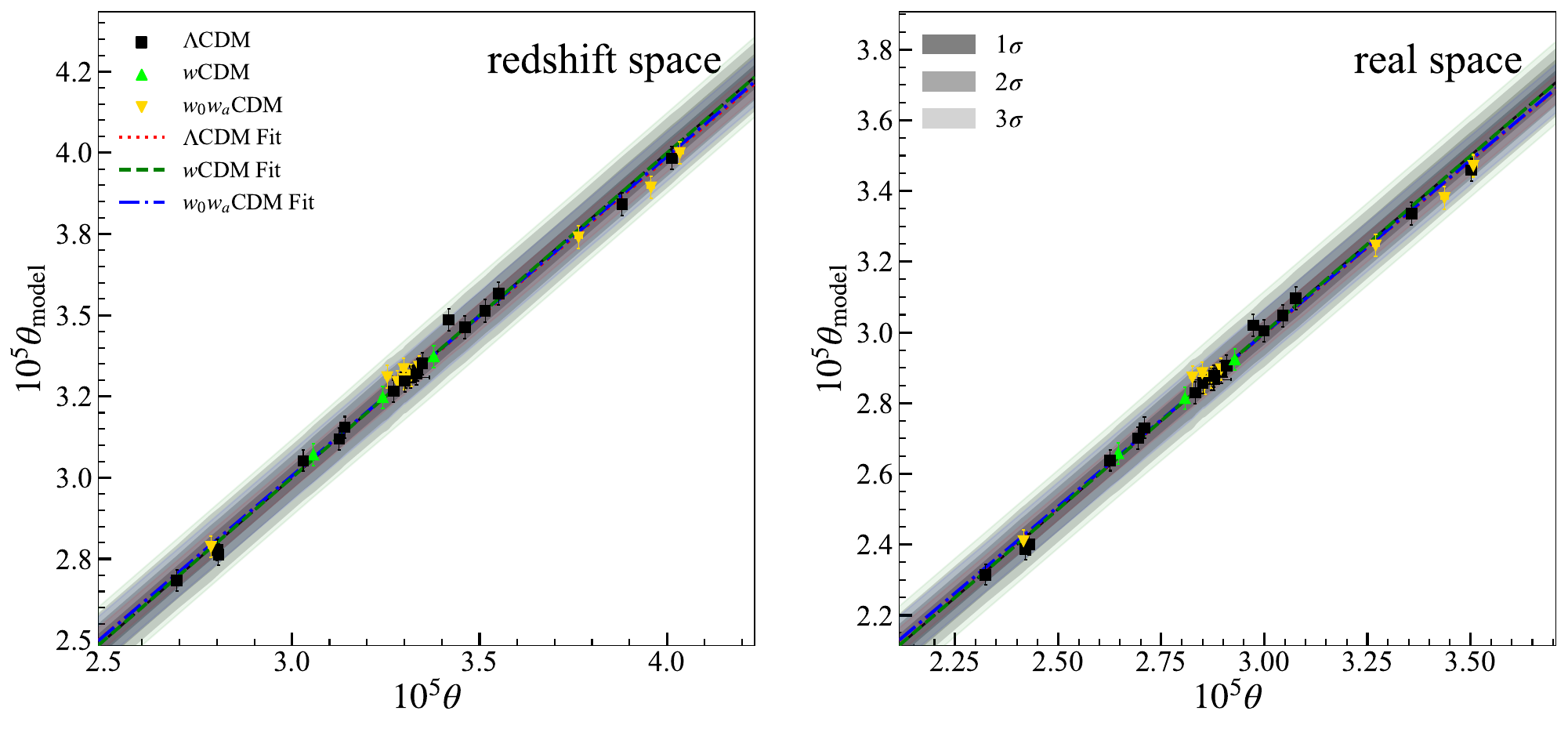}
      \caption{Redshift and real space configurations of the three classes of the background cosmologies in $\theta$-$\theta_{\rm model}$ plane. 
      In each panel, the shaded (dark, light and lightest) gray-color area correspond to the $68\%$, $95\%$ and $99\%$ 
confidence regions around the best-fit $\theta_{\rm model}$ (red dotted line) determined for the $\Lambda$CDM cosmologies. For comparison, the 
same bilinear models, $\theta_{\rm model}$, but with its coefficients obtained for the $w$CDM (blue dashed line) and $w_{0}w_{a}$CDM cosmologies are 
displayed (blue dashed and navy dot-dashed lines, respectively) in each panel. }
         \label{fig:t_m}
   \end{figure}

Given the results shown in Fig.~\ref{fig:t_fix}, we put forth the following non-parametric bilinear model for $\theta$:
\begin{equation}
\label{eqn:t_m}
10^{5}\theta_{\rm model}(\Omega_{m},\sigma_{8})=\eta_{1}\Omega_{m}+\eta_{2}\sigma_{8}+\eta_{3}\, ,
\end{equation}
where $\{\eta_{i}\}_{i=1}^{3}$ are three coefficients whose values are determined via the generalized $\chi^{2}$-statistics for the $\Lambda$CDM case 
(the second row of Table~\ref{tab:eta_lcdm}). 
The left-panel of Fig.~\ref{fig:t_m} plots $\theta_{\rm model}$ with these best-fit coefficients versus the original scale parameter $\theta$ 
as black filled squares for the $\Lambda$CDM cosmologies. The shaded dark, light, and lightest gray regions around the straight line of 
$\theta=\theta_{\rm model}$ correspond to the $1\sigma$, $2\sigma$ and $3\sigma$ scatters around $\theta$. 
As can be seen,  Eq.~(\ref{eqn:t_m}) matches very well the original values of $\theta$, for the case of the $\Lambda$CDM cosmologies. 
\begin{table}[ht!]
\caption{Best fit coefficients of $\theta_{\rm model}(\Omega_{m},\sigma_{8})$ obtained for the case of the $\Lambda$CDM cosmologies.}  
\label{tab:eta_lcdm}    
\centering                        
\begin{tabular}{l c c c}      
\hline\hline  \\[-0.5em]             
space & $\eta_{1}$ & $\eta_{2}$ & $\eta_{3}$ \\   [\smallskipamount]      
\hline\\[-0.4em]
   redshift & $9.12 \pm 0.03$ & $2.72\pm 0.02$ &  $-1.69\pm 0.02$ \\   [\smallskipamount] 
   real & $8.1\pm 0.03$ & $2.43\pm 0.02$    & $-1.58 \pm 0.02$ \\ [\smallskipamount] 
\hline
\end{tabular}
\end{table}

For the $w$CDM and $w_{0}w_{a}$CDM cosmologies, we evaluate $\theta_{\rm model}$ by putting their values of $\Omega_{m}$ and 
$\sigma_{8}$ into Eq.~(\ref{eqn:t_m}), while setting the values of $\{\eta_{i}\}_{i=1}^{3}$ at the same best-fit values obtained 
for the $\Lambda$CDM case.  In the left panel of Fig.~\ref{fig:t_m} are plotted $(\theta, \theta_{\rm model})$ as filled triangles for these two classes. 
Note that $(\theta,\theta_{\rm model})$ are all within $3\sigma$ from the linear line of $\theta=\theta_{\rm model}$, for the $w$CDM and $w_{0}w_{a}$CDM cases, 
even though the coefficients, $\{\eta_{i}\}_{i=1}^{3}$, in Eq.~(\ref{eqn:t_m}) are not adjusted but set at the $\Lambda$CDM values. 
This result implies that Eq.~(\ref{eqn:t_m}) with constant coefficients are universally valid for all of the three cosmology classes considered. 

To confirm further the universality of our bilinear model, we re-determine the coefficient values of our bilinear model by fitting Eq.(\ref{eqn:t_m}) to $\theta$ 
determined for the $w$CDM and $w_{0}w_{a}$CDM cases (blue long-dashed and navy dot-dashed lines, respectively, in the left panel of Fig.\ref{fig:t_m}). 
As can be seen, only negligibly small difference exist from the $\Lambda$CDM case (red dotted line). This result indicates that no matter what cosmology class 
is used to determine its coefficients, $\{\eta_{i}\}_{i=1}^{3}$, our bilinear model, Eq.(\ref{eqn:t_m}), validly describes how $\theta$ changes 
with each of $\Omega_{m}$ and $\sigma_{8}$, explaining why the best-fit coefficients of $\theta_{\rm model}$ obtained for the $\Lambda$CDM case work 
even for the other two classes.

To examine whether or not the void asymmetry distributions determined in real-space would yield a similar result, we repeat the 
whole process but in real space. Basically, the same mass cutoff for the galactic halos, the same Void-Finder algorithm,  the same 
number-cutoff for the void member galaxies, and the same iterative routine are employed for the determinations of $p(\Delta z_{\rm max})$ as well as its parameter 
$\theta$ in real space.  An almost perfect linear dependence between $\theta$ and each of $\Omega_{m}$ and $\sigma_{8}$ is also witnessed in real space 
(bottom panels of Fig.\ref{fig:t_fix}). Eq.(\ref{eqn:gam}) is also found to describe well the real space $p(\Delta z_{\rm max})$ (right panels of Fig.\ref{fig:gam}). 
The real space bilinear model, $\theta_{\rm model}$, is also determined by fitting the real space $\theta$ values to Eq.(\ref{eqn:t_m}) and determining 
the coefficient values (the third row of Table \ref{tab:eta_lcdm}). Eq.(\ref{eqn:t_m}) with real space coefficients is also found to match very well the real space 
best-fit $\theta$ values (right panel of Fig.\ref{fig:t_m}). As can be read in Table \ref{tab:eta_lcdm}, however, all of the 
three coefficients, $\{\eta_{i}\}_{i=1}^{3}$, have substantially lower best-fit values in real space than in redshift space. In other words, the void asymmetry 
distribution $p(\Delta z_{\rm max})$ has a lower scale parameter in real space than in redshift space. 
Given that $\theta$ quantifies the spread of a Gamma distribution and that the redshift space distortion plays a role of causing 
additional uncertainties in the measurement of $\Delta z_{\rm max}$, this result reflects that the void redshift asymmetries, if measured in redshift space, 
are subject to larger uncertainties, and thus described by more dispersed distributions.

   \begin{figure}[ht!]
   \centering
   \includegraphics[width=\hsize]{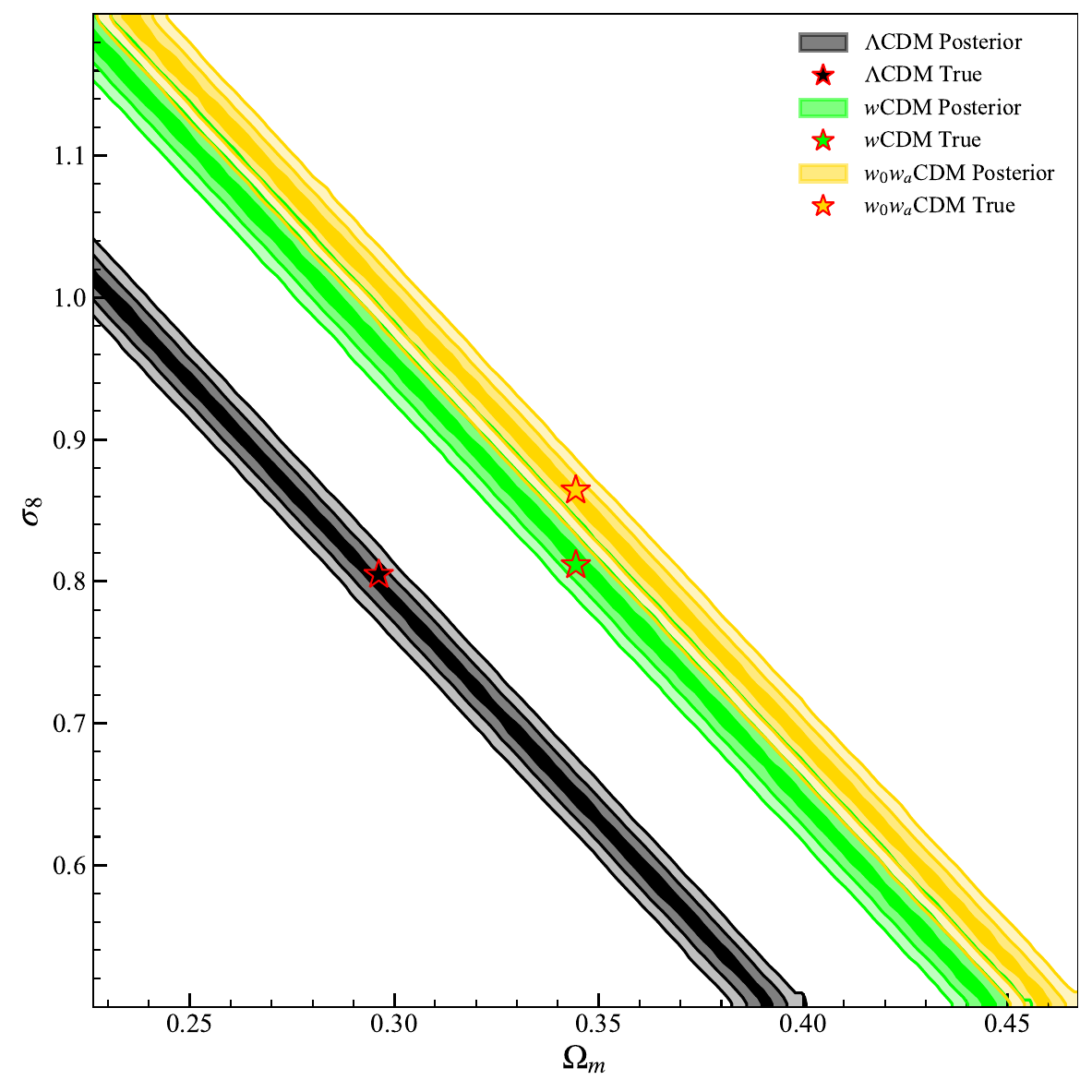}
      \caption{$68\%$, $95\%$ and $99\%$ contours obtained by fitting $\theta_{\rm model}$ given in Eq.(\ref{eqn:t_m}) to $\theta$ by adjusting $\Omega_{m}$ and $\sigma_{8}$ for three 
      additional cosmologies: one $\Lambda$CDM (gray contours), one $w$CDM (green contours) and one $w_{0}w_{a}$CDM (yellow contours) cosmologies. These three additional cosmologies have been 
      chosen as the different ones that are not included in the original sets of the $33$ $\Lambda$CDM cosmologies. 
      The actual values of $\Omega_{m}$ and $\sigma_{8}$ of these three cosmologies are marked as open red star symbols.}
         \label{fig:robust1}
   \end{figure}
   \begin{figure}[ht!]
   \centering
   \includegraphics[width=\hsize]{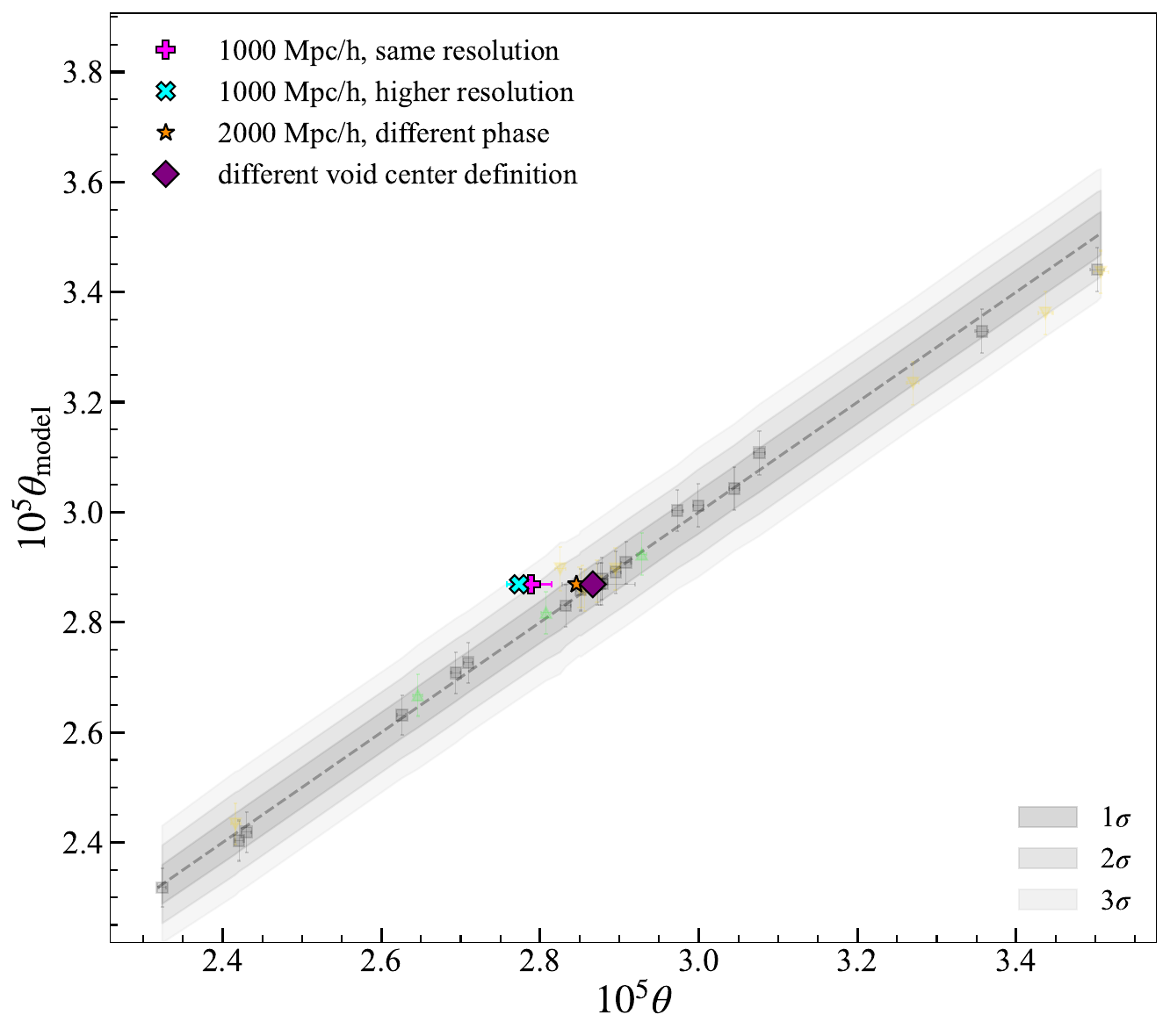}
      \caption{Locations of $(\theta,\theta_{\rm model})$ determined in real space from four different cases: 
      One corresponds to the case that the void centers are misidentified (purple diamond symbols), and the rest three cases are from three additional runs of the AbacusSummit simulations 
      that are different in the simulation volume  (magenta plus symbol), particle mass resolution (cyan cross symbol) and phase (orange star symbol) from the original run used to 
      determine the coefficients of $\theta_{\rm model}$. 
      The shaded (dark, light and lightest) gray-color area correspond to the $68\%$, $95\%$ and $99\%$ confidence regions around the original linear $\theta$-$\theta_{\rm model}$ relation 
      shown in Fig.\ref{fig:t_m}.}
         \label{fig:robust2}
   \end{figure}

 \section{Algorithm feasibility}

\subsection{Constraining $\sigma_{8}$ and $\Omega_{m}$ with $\theta$}\label{sec:feasibility}

In practice, one can take the followings steps to put complementary constraints on $\Omega_{m}$ and $\sigma_{8}$ from the observed void asymmetry 
distribution in redshift space. 
\begin{itemize}
\item
Identify voids in the local universe via the Void-Finder algorithm and select only those voids containing $15$ or more member galaxies beyond 
the flux limit in redshift space. 
\item
Estimate the size distribution of selected voids, $p(R_{v})$, and then control $p(R_{v})$ to have the same shape as the numerical result shown in 
the right panel of Fig.\ref{fig:nrv}. 
\item
Measure the redshift asymmetry between the bisected parts of each void by iteratively altering the orientation of a bisector line until it converges to a maximum value, $\Delta z_{\rm max}$. 
\item
Determine the probability density function, $p(\Delta z_{\rm max})$, from the selected voids.
\item
Compare the generalized gamma distribution with fixed $k=6.551$ to the observed $p(\Delta z_{\rm max})$ by adjusting $\theta$ via the 
maximum likelihood estimation.
\item
Compare the best-fit value of $\theta$ to $\theta_{\rm model}$ given in Eq.(\ref{eqn:t_m}) by adjusting $\Omega_{m}$ and $\sigma_{8}$ via 
the maximum likelihood estimation. Here, $\theta_{\rm model}$ is computed by plugging into Eq.~(\ref{eqn:t_m}) the universal coefficient values given in the first row 
of Table~\ref{tab:eta_lcdm} that have been obtained for the $\Lambda$CDM cosmology class.
\item 
Compute the significance contours that constrain the 2D parameter space spanned by $\Omega_{m}$ and $\sigma_{8}$. 
\end{itemize}

We test this algorithm against three different datasets, each of which is extracted from the same AbacusSummit suite but performed for three different cosmologies that 
have not been included in the original sets of $33$ cosmologies.  Each of these three additional cosmologies belongs to the $\Lambda$CDM, $w$CDM and $w_{0}w_{a}$ classes, respectively. 
Applying this algorithm to these datasets, we obtain the posterior contours in the $\Omega_{m}$-$\sigma_{8}$ plane, which is displayed in Fig.\ref{fig:robust1}. 
The actual values of ($\Omega_{m}$, $\sigma_{8}$) (open red stars) are found to be well within the $68\%$ significance contours for all of the three cases, validating 
the feasibility of this algorithm. 

\subsection{Robustness tests}\label{sec:robustness}

To test the robustness of our bilinear model, we first investigate whether or not possible mis-identification of void centers would affect the final results.  
As mentioned in section \ref{sec:analysis}, the Void-Finder identifies a void as an assembly of several overlapping empty spheres, among which the main empty sphere is the one that 
has the largest radius. Instead of using the true center of each void that coincides with the centroid of void galaxies,  we use the center of its main empty sphere and repeat the whole analysis 
for the case of the Planck $\Lambda$CDM cosmology.  As the redshift-space distortion effects would play a role of somewhat obscuring the errors caused by mis-identification 
of a void center by increasing the scatters in the measurement of void redshift asymmetries, we perform this robustness test in real space rather than in redshift space.
The best-fit value $\theta$ is re-determined and compared with $\theta_{\rm model}$ given in Eq.(\ref{eqn:t_m}) with the same coefficients 
given in the second row of Table~\ref{tab:eta_lcdm}, the result of which is shown as a purple diamond symbol in Fig.~\ref{fig:robust2}.  
As can be seen, the misidentification of void centers has a negligible effect on the final result. 

The robustness of our bilinear model, $\theta_{\rm model}(\sigma_{8},\Omega_{m})$, is also tested against the variation of simulation.  
For this test, we utilize three other runs of the AbacusSummit simulation series, different from the original runs used in the current analysis to derive the coefficient values of Eq.(\ref{eqn:t_m}). 
One was performed on a smaller volume of $1\,h^{-3}{\rm Gpc}^{3}$ but with the same particle mass resolution as the original run, the other was performed on a smaller volume 
$1\,h^{-3}{\rm Gpc}^{3}$ with a higher particle mass resolution of $m_{p}=3.5\times 10^{8}\,h^{-1}\,M_{\odot}$, while another is on the same volume with the same $m_{p}$ but 
from a different phase.  All of the three different runs assumed the Planck $\Lambda$CDM cosmology. 
Again, to avoid any obscuration caused by redshift space distortion effects and to single out the effects of variations of simulation phase and resolution on the measurements of void 
asymmetry distributions, we perform this second robustness test in real space, too.

Repeating the same process with a dataset from each of these three different runs, we determine $p(\Delta z_{\rm max})$ and its best-fit parameter $\theta$. Putting the actual 
values of $\Omega_{m}$ and $\sigma_{8}$ of each run into Eq.(\ref{eqn:t_m}), we also compute $\theta_{\rm model}$ under the assumption that the same bilinear model should be valid. 
Fig.~\ref{fig:robust2} plots $(\theta,\theta_{\rm model})$ from the three different simulation runs (magenta plus, cyon cross and orange star symbols). 
As can be seen,  all of the three points are located within the $99\%$ confidence regions from the original linear $\theta$-$\theta_{\rm model}$ relation. This result verifies the robustness of 
our bilinear model against the variation of simulation volume, mass resolution and phase. 
Note, however, that the largest deviation from the original linear relation is witnessed for the case that $\theta$ is obtained from the simulation of highest resolution, which implies 
that the best-fit  coefficients of our bilinear model are resolution-limited approximations. 

\section{Summary and conclusions}\label{sec:con}

In light of our prior result that the void spin distributions exhibit exclusive dependence on $(\Omega_{m},\sigma_{8})$ regardless of $w$~\citep{2025JCAP...06..011K}, 
the current work has numerically explored an observational proxy of the void spin distributions satisfying the following two conditions: 
First, this proxy should be expressed in terms of readily measurable quantities defined in redshift space without requiring any information on the peculiar velocities of void galaxies. 
Second, it should retain the same or at least similar degree of sensitivity to $(\Omega_{m},\sigma_{8})$ and independence on $w$. This exploration, based on the mock redshift space 
catalog of galactic halos at $z=0.1$ from the Abacus-Summit suite of simulations with various initial conditions~\citep{2021MNRAS.508.4017M}, has led us to discover 
that the void asymmetry distribution is indeed a good proxy of the void spin distribution, fully satisfying both of the aforementioned two conditions. 

In the current work has been considered a total of $33$ different ($20$ $\Lambda$CDM, $3$ $w$CDM and $10$ $w_{0}w_{a}$CDM) cosmologies.  
For each cosmology, the redshift space voids have been identified from the mock galaxy catalog via the classical Void-Finder algorithm~\citep{2002ApJ...566..641H} 
to be consistent with \citet{2025JCAP...06..011K}. 
Employing the iterative algorithm developed by \citet{2021NatAs...5..839W} for the determination of the spin axes of large-scale structures (LSS), we have searched for 
the spin axis of each void in the 2D plane of sky, which bisects it into two parts between which the redshifts of void member galaxies relative to its center exhibit the 
maximum degree of asymmetry. The probability density functions of the maximum redshift asymmetries of the void galaxies are then computed as the void asymmetry distributions  
from the size-controlled sample of voids for each cosmology. 

Just like the void spin distributions~\citep{2025JCAP...06..011K}, the void asymmetry distribution have been found to be excellently described by the generalized Gamma model 
characterized by the shape and scale parameters. We have for the first time shown that while the shape parameter, $k$, has turned out to be almost independent of the background cosmology, 
the scale parameter, $\theta$, exhibits a strong linear dependence on each of $\Omega_{m}$ and $\sigma_{8}$, regardless of $w$ values.
We have also constructed a new non-parametric bilinear model for $\theta$ as a function of $\Omega_{m}$ and $\sigma_{8}$ and demonstrated that this model, $\theta_{\rm model}$, is 
valid for all of the $\Lambda$CDM, $w$CDM and $w_{0}w_{a}$CDM cosmologies considered, even though its three coefficients are fixed at the best-fit values obtained 
for the case of the $\Lambda$CDM cosmologies. We have also confirmed that the detected linear dependence of $\theta_{\rm model}$ on each of $\Omega_{m}$ and $\sigma_{8}$ are not 
just secondary effects caused by any differences in void size distributions among background cosmologies nor by redshift space distortion effects. 
Given these results, we have proposed the void asymmetry distribution as a new complementary diagnostics, which can in principle break the 
$\Omega_{m}-\sigma_{8}$ degeneracy, if combined with the standard LSS-based diagnostics like the weak lensing statistics that can determine $S_{8}\equiv \sigma_{8}\left(\Omega_{m}/0.3\right)^{0.5}$~\citep[e.g.,][]{2013MNRAS.430.2200K,2022PhRvX..12c1029K}. 

A feasibility of this new diagnostics of ($\Omega_{m}$, $\sigma_{8}$) based on our bilinear model for the scale parameter of void asymmetry distribution has been tested against three 
additional simulations from the AbacusSummit suite that are not included for the determination of the model coefficients. It has been shown that the actual values of $\Omega_{m}$ and 
$\sigma_{8}$ of these three extra cosmologies are located well within the $1\sigma$ confidence regions constrained by $\theta_{\rm model}(\Omega_{m},\sigma_{8})$. 
Yet, testing the robustness of $\theta_{\rm model}(\Omega_{m},\sigma_{8})$ against the variations of simulation volume,  particle mass resolution and phase as well as possible 
mis-identification of void centers, we have shown that although our bilinear model is quite universal and robust, it will be desirable to update its coefficients by utilizing higher-resolution 
simulations for the improvements of its efficiency and validity.

The void asymmetry distribution has a couple of advantages  as an independent diagnostics of $(\Omega_{m},\sigma_{8})$ over the other LSS probes.
First, it requires information only on the redshifts of void galaxies in the local universe, the measurements of which have achieved the 
unprecedentedly high $1\%$ accuracy level in the currently available data from the latest galaxy surveys~\citep{2015ApJS..219...12A}.  
Second, it is almost independent of $w$ unlike the standard LSS-based probes like the cosmic shear statistics, redshift space distortion effects 
and evolution of the cluster abundance~\citep[e.g.,][]{1998ApJ...508..483W,2008Natur.451..541G,2015RPPh...78h6901K}, 
all of which are intricately degenerate among $\Omega_{m}$, $\sigma_{8}$ and $w$. 
Third, unlike the other LSS-based diagnostics, it is relatively free from astrophysical uncertainties~\citep{2022PhRvD.105b3515S}, given that the voids are the most pristine structures 
whose evolutions are expected to be least complicated by the late-time nonlinear astrophysical processes. 

It is, however, worth mentioning possible uncertainties associated with redshift errors, $\Delta z \sim 0.02-0.03$, that the photometric galaxy catalog from the 
LSST, our reference survey, has been known to carry \citep{2009arXiv0912.0201L}. In practice, these observational redshift errors would affect the identification of voids and determination of 
their asymmetry distributions, which in turn would modify the coefficient values of our bilinear model. Thus, it will be definitely necessary to perform a more thorough analysis based on mock 
galaxy catalogs from light-cone simulations that mimic the LSST photometric galaxy catalog  and to explore the effects of observational redshift errors on the efficiency 
and power of void asymmetry distributions as a probe of $\Omega_{m}$ and $\sigma_{8}$.
Our future work will be in the direction of conducting these additional numerical investigations and then measuring the void asymmetry distribution from real observational data to 
tightly constrain the values of $\Omega_{m}$ and $\sigma_{8}$ in combination with the standard cosmological diagnostics.

\begin{acknowledgements}
G.K. acknowledges that this research was supported by Basic Science Research Program through the National Research Foundation of Korea (NRF) 
funded by the Ministry of Education (RS-2025-25408975).
J.L. acknowledges the support by Basic Science Research Program through the  NRF of Korea funded by the Ministry of Education (RS-2025-00512997). 
\end{acknowledgements}

\bibliographystyle{aa.bst} 
\bibliography{ref.bib}

\end{document}